\documentclass[aip, jmp, prb, amsmath, amssymb, reprint]{revtex4-1}

\usepackage{csquotes}
\usepackage[pdftex]{graphicx}
\usepackage{epstopdf}

\usepackage[table]{xcolor}

\usepackage[section]{placeins}

\usepackage{booktabs}
\usepackage{array}

\usepackage[euler]{textgreek}
\usepackage{bm}

\usepackage[breaklinks=true,colorlinks=true,linkcolor=blue,urlcolor=blue,citecolor=blue]{hyperref}

\usepackage{amssymb}
\usepackage{amsmath}

\bibpunct{[}{]}{,}{n}{}{} 

\usepackage{subcaption}
\newcommand{\labelphantom}[1]{%
{\phantomsubcaption%
\label{#1}}%
}%
\makeatletter
\renewcommand\p@subfigure{\thefigure-}
\makeatother

\begin{document}

\title[Assessment of vertical stability for negative triangularity pilot plants]{Assessment of vertical stability for negative triangularity pilot plants} 

\author{S. Guizzo}
\affiliation{Columbia University, New York City, New York 10027, USA}

\author{A.O. Nelson} 
\email[Corresponding author: ]{a.o.nelson@columbia.edu}
\affiliation{Columbia University, New York City, New York 10027, USA}

\author{C. Hansen} 
\affiliation{Columbia University, New York City, New York 10027, USA}

\author{F. Logak} 
\affiliation{École Polytechnique, Paris, France
}

\author{C. Paz-Soldan}
\affiliation{Columbia University, New York City, New York 10027, USA}

\vspace{10pt}
\date{\today}

\begin{abstract}

Negative triangularity (NT) tokamak configurations may be more susceptible to magneto-hydrodynamic instability, posing challenges for recent reactor designs centered around their favorable properties, such as improved confinement and operation free of edge-localized modes. In this work, we assess the vertical stability of plasmas with NT shaping and develop potential reactor solutions. When coupled with a conformal wall, NT equilibria are confirmed to be less vertically stable than equivalent positive triangularity (PT) configurations. Unlike PT, their vertical stability is degraded at higher poloidal beta. Furthermore, improvements in vertical stability at low aspect ratio do not translate to the NT geometry. NT equilibria are stabilized in PT vacuum vessels due to the increased proximity of the plasma and the wall on the outboard side, but this scenario is found to be undesirable due to reduced vertical gaps which give less spatial margin for control recovery.  Instead, we demonstrate that informed positioning of passively conducting plates can lead to improved vertical stability in NT configurations on par with stability metrics expected in PT scenarios. An optimal setup for passive plates in highly elongated NT devices is presented, where plates on the outboard side of the device reduce vertical instability growth rates to 16\% of their baseline value. For lower target elongations, integration of passive stabilizers with divertor concepts can lead to significant improvements in vertical stability. Plates on the inboard side of the device are also uniquely enabled in NT geometries, providing opportunity for spatial separation of vertical stability coils and passive stabilizers.

\end{abstract}

\vspace{2pc}
\keywords{negative triangularity, vertical stability, VDE}

\maketitle


\section{Introduction}
\label{sec:intro}
 Tokamak scenarios with negative triangularity (NT) shaping ($\delta <0$) are a topic of recent interest for reactor design due to promising results from TCV \cite{pochelon_energy_1999, Camenen2007}, DIII-D \cite{ Austin2019, marinoni_h-mode_2019, marinoni_diverted_2021}, and AUG \cite{merle_pedestal_2017} experiments.  NT plasmas have been shown to exhibit impressive core performance, with DIII-D discharges reaching $H_{98,y2} \geq 1$, $\beta_{N} \sim 3$, and $f_{GW}>1$ at $q_{95} \leq 3$ \cite{paz_soldan_2023}, and gyrokinetic predictions indicate that favorable confinement properties will persist in larger devices \cite{merlo_effect_2023}.  Furthermore, NT configurations achieve this confinement while maintaining a reduced edge pressure gradient and operation free of edge-localized modes (ELMs) \cite{nelson_prospects_2022,saarelma_ballooning_2021,merle_pedestal_2017, nelson_robust_2023}. The NT geometry also places strike points on the outboard, low-field side of the device, resulting in an inherently larger surface area for heat deposition and reduced heat flux on divertor plates \cite{Medvedev2015}. Together, these characteristics make negative triangularity a promising configuration for a future fusion pilot plant which will require robust power handling and ELM-free operation in tandem with a high-performance, fusion-relevant core plasma \cite{Kikuchi2019, Medvedev2015, frank_radiative_2022}.

However, NT plasmas are also generally subject to degraded magneto-hydrodynamic (MHD) stability \cite{weisen_effect_1997, xue_hot_2019, ren_comparative_2016, zheng_intermediate_2021}, warranting further study of the configuration. In particular, NT shaping is predicted to increase susceptibility to the $n=0$ vertical instability, which could place an upper-bound on the plasma elongations accessible in an NT reactor.  Because energy confinement time scales with elongation ($\tau_E \propto \kappa^{0.7}$) \cite{iter_physics_expert_group_on_confinement_and_transport_chapter_1999}, this constraint could limit the maximum fusion power accessible by NT scenarios.  Furthermore, poor intrinsic vertical stability increases the likelihood of vertical displacement events (VDEs), which trigger major disruptions that must be avoided in future machines.

A vertical control system can prevent a VDE when the instability growth rate is sufficiently small and the coil current responses can diffuse through the vacuum vessel in sufficiently short times.  The feedback capability parameter, which is defined as the product of the instability growth rate ($\gamma$) and wall diffusion time ($\tau_{\mathrm{w}}$) can be used to characterize the robustness of a given control system \cite{freidberg_tokamak_2015}. Some current machines such as the DIII-D tokamak can control $\gamma \tau_{W}\lesssim 6$ \cite{lee_tokamak_2015}. However, $\gamma \tau_{\mathrm{w}}\lesssim1.5$ is a reasonable estimate of what is expected to be controllable in future fusion-grade experiments, where feedback coils will need to be located farther from the plasma outside of toroidal field coils in order to be shielded from neutron radiation \cite{lee_tokamak_2015}.  

This constraint on active control necessitates investigation of the passive vertical stability of reactor scenarios, particularly for relatively unstudied configurations like NT. Extensive modeling with the \texttt{AVSTAB} (Axisymmetric Vertical STABility) code \cite{saarelma_ballooning_2021} has laid the groundwork for understanding vertical stability in NT configurations, which will be extended in this study.  \texttt{AVSTAB} results demonstrate that NT plasma shapes in conformal wall geometries have lower marginally controllable elongations than equivalent positive triangularity (PT) plasmas, potentially because NT equilibria have more elongated inner flux surfaces \cite{song_impact_2021}. 

\vspace{12pt}
Interestingly, NT vertical stability seems to be particularly sensitive to the shape of the conducting vacuum vessel. Modeling with \texttt{AVSTAB} showed that greater marginally controllable elongations can be achieved when NT equilibria are coupled with a positive triangularity vacuum vessel due to the decreased distance between the outboard corners of the plasma and the conducting wall \cite{song_impact_2021}.  
Similarly, simulations of NT configurations on the DIII-D tokamak analyzed in \texttt{TokSys} \cite{humphreys_development_2007} revealed that more negative triangularity does not always correspond to larger instability growth rates in the presence of a conducting wall \cite{Nelson2023}. Instead, when triangularity decreased sufficiently, the outboard corners of the NT plasma shape became increasingly close to the positive triangularity vacuum vessel, resulting in a net stabilizing effect \cite{Nelson2023}. These observations suggest that NT vertical stability has the potential to be improved through strategic placement of stabilizing conductors.

In this study, we determine the dependence of vertical instability growth rates on plasma and machine properties in order to gauge the feasibility of favorable passive stability in NT reactor scenarios.  In section~\ref{sec:pla},  we begin by using the \texttt{TokaMaker} code \cite{hansen_2023} to assess various vertical stability trends for free-boundary NT equilibria in conformal wall geometries. Vacuum vessel shape is discussed in section~\ref{sec:nonc}, where we find that reduced growth rates in non-conformal wall geometries do not compensate for decreased vertical gaps in compact reactors, and are therefore not the ideal resolution to the NT vertical stability question. However, in section~\ref{sec:plate} we determine the most impactful locations for passive stabilizing plates in NT geometries and demonstrate that NT vertical stability issues can be largely mitigated via optimization of passive conducting structures. Finally, conclusions and outlooks for future machines are presented in section~\ref{sec:conc}.

\section{Effect of plasma geometry and equilibrium profiles}
\label{sec:pla}
To investigate the effect of plasma properties and conducting structures on vertical stability, the open-source \texttt{TokaMaker} code \cite{hansen_2023} is used to generate plasma equilibria over a large parameter space. \texttt{TokaMaker} is a new 2D free-boundary Grad-Shafranov equilibrium solver with both static and time dependent capabilities. \texttt{TokaMaker} can also compute the most unstable eigenmodes of the linearized system along with their corresponding growth rates and the diffusion times for conducting structures.  By identifying the $n=0$ system and wall eigenmodes, one can calculate $\gamma \tau_{\mathrm{w}}$ for various scenarios. A benchmark of \texttt{TokaMaker} against the well-established \texttt{TokSys} code \cite{humphreys_development_2007} is presented in Appendix \hyperref[sec:val]{A}.

The baseline device used in this study has target parameters similar to the ARC \cite{sorbom_arc_2015} and MANTA \cite{manta} tokamaks, including  plasma major radius $R_0 = 3.3\,$m, minor radius  $a = 1.1\,$m, plasma current $I_p = 8\,$MA, and on-axis magnetic field $B_0 = 10\,$T.  All equilibria considered are diverted and up-down symmetric.  The pressure ($p$) and radially scaled azimuthal magnetic field ($F$)  are assumed to have fixed profiles of the form

\begin{equation}
    \frac{dp}{d\psi_\mathrm{N}} = p_0(1-\psi_\mathrm{N}^2)^2
    \label{eq:pres}
\end{equation}

\begin{equation}
    \frac{1}{2}\frac{dF^2}{d\psi_\mathrm{N}} = f_0(1-\psi_\mathrm{N}^2)^2
    \label{eq:f}
\end{equation}
where $\psi_\mathrm{N}$ is the normalized poloidal magnetic flux and $p_0$ and $f_0$ are constants corresponding to the on-axis pressure gradient and azimuthal magnetic flux gradient which are chosen to achieve a target poloidal beta $\beta_p = 1$. The profile shape corresponds to internal inductance $l_\mathrm{i} \approx 1$. Elongation of $\kappa = 1.8$ is chosen as a target to test the limits of vertical controllability. This baseline scenario is modified by scanning the quantity of interest, while keeping all other parameters fixed.

\begin{figure}
	\includegraphics[width=1\linewidth]{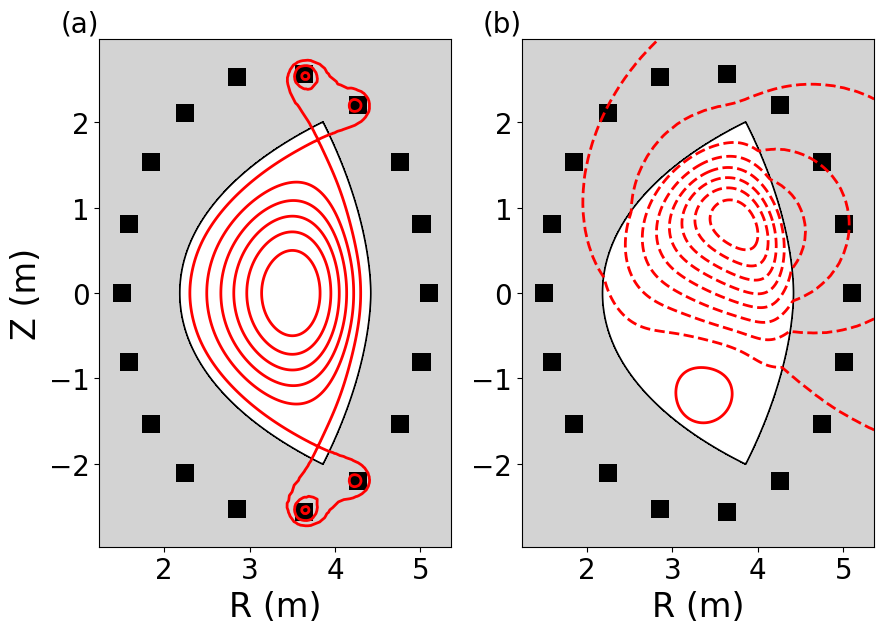}
	\caption{a) Sample equilibrium flux and b) perturbed flux of vertically unstable eigenmode for $\delta = -0.5$}
	\label{fig:cartoon}
\end{figure}

Unlike previous NT vertical stability investigations, we use free-boundary equilibria generated with strict constraints on the target plasma boundary.  
It is therefore necessary to define a poloidal field (PF) coil set that can generate all desired boundary shapes to high accuracy, even as geometric properties such as triangularity and elongation are varied.  A fixed PF coil set of 18 coils evenly spaced around an ellipse surrounding the vacuum region is always sufficient to generate the desired target shape.  Although the chosen PF coil set impacts the vacuum magnetic field and thus the absolute magnitude of the vertical instability growth rates, using fixed coil locations for all configurations ensures that the coil set does not impact the observed qualitative trends.

Throughout this section, a conformal wall is assumed in order to isolate the effects of the plasma geometry and equilibrium profiles from that of the passive conducting structures, which are considered in sections~\ref{sec:nonc} and ~\ref{sec:plate}. The wall is placed at $b/a = 1.1$, where $b$ is the minor radius of the vacuum vessel. For reference, figure~\ref{fig:cartoon} shows a sample device and equilibrium generated in \texttt{TokaMaker} along with the perturbed flux of the largest vertically unstable eigenmode. The PF coil set used to calculate free-boundary equilibria is also depicted, emphasising flexibility in plasma shape control.


\subsection{Triangularity and Elongation}



\begin{figure}[h]
	\includegraphics[width=1\linewidth]{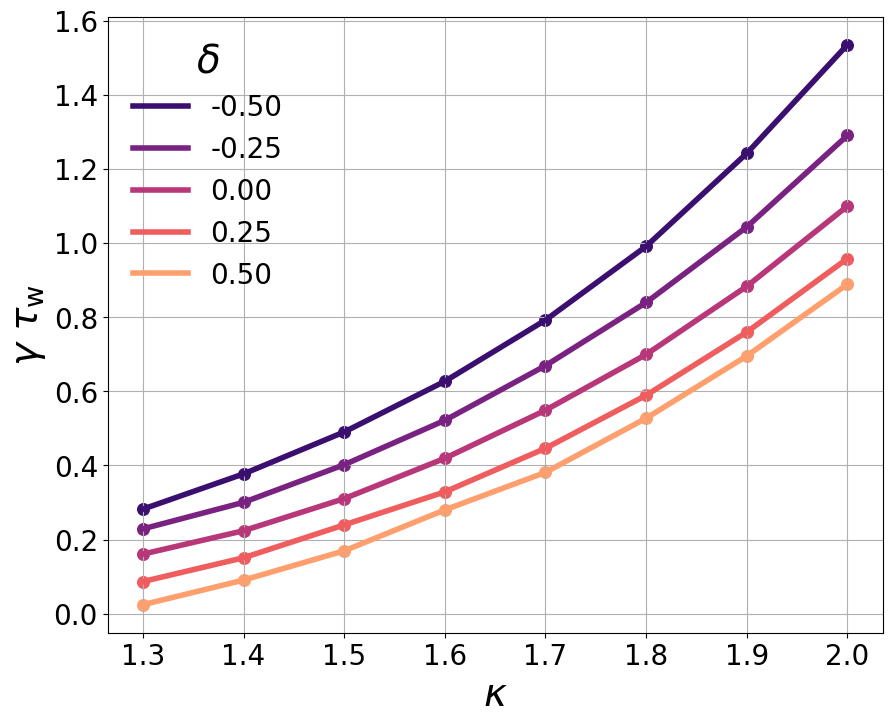}
	\caption{Feedback capability parameter $\gamma \tau_w$ as a function of $\kappa$ for equilibria with different $\delta$.}
	\label{fig:delta}
\end{figure}

The scenario described above serves as a starting point to explore the dependence of vertical stability on equilibrium properties. As shown in figure~\ref{fig:delta}, feedback capability parameter calculations in \texttt{TokaMaker} show that $\gamma \tau_{\mathrm{w}}$ increases with increasing elongation and decreasing triangularity, as expected. NT equilibria are found to have more elongated inner flux surfaces when compared to positive triangularity, which is a fundamental consequence of their geometry.  Previous studies with \texttt{AVSTAB} indicate that increased elongation of inner flux surfaces resultes in more negative $\delta W_{\mathrm{fluid}}$ contributions to the total $\delta W$ Lagrangian integral \cite{song_impact_2021} for NT configurations, which is consistent with larger instability growth rates observed for the free-boundary equilibria generated in \texttt{TokaMaker}. 

Inferior vertical stability means that NT plasmas are limited to lower values of elongation than PT plasmas to stay below stability thresholds, which is undesirable for a fusion pilot plant due to the positive correlation between energy confinement time and plasma elongation \cite{iter_physics_expert_group_on_confinement_and_transport_chapter_1999}.  However, as demonstrated below in sections~\ref{sec:nonc} and ~\ref{sec:plate}, placement of stabilizing conductors can lead to improvements in NT vertical stability.

\subsection{Equilibrium Profiles}


    

\begin{figure*}
   \labelphantom{fig:profiles-a}
   \labelphantom{fig:profiles-b}
   \labelphantom{fig:profiles-c}
    \centering
    \includegraphics[width=1\linewidth]{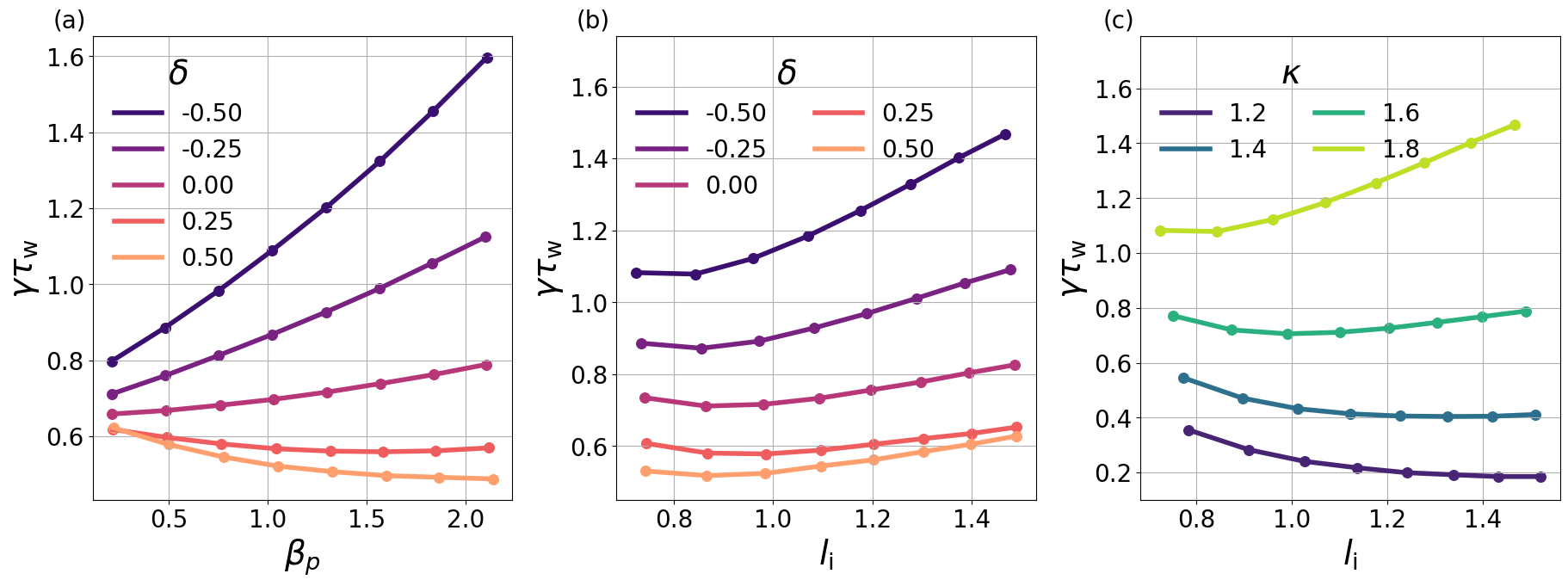} %
    \caption{Feedback capability parameter $\gamma \tau_w$ as a function of (a) $\beta_{p}$ for equilibria with different $\delta$, (b) $\l_\mathrm{i}$ for equilibria with different $\delta$ and (c) $\l_\mathrm{i}$ for equilibria with different $\kappa$.} %

    \label{fig:prof}%
    
\end{figure*}

The geometry of internal flux surfaces, which contributes to the relatively degraded vertical stability observed in NT configurations, depends on the equilibrium current and pressure profiles. Current and pressure profiles are also important in dictating the strength of wall stabilization by changing the effective distance between the plasma and conducting vacuum vessel. The interplay between these two effects depends on the plasma triangularity and elongation in question, and is therefore a topic of interest for NT vertical stability.

NT configurations are less vertically stable at higher values of $\beta_{p}$, as shown in figure~\ref{fig:profiles-a}.  Equilibria with negative $\delta$ exhibit greater elongation of inner flux surfaces at higher $\beta_{p}$, which is consistent with larger instability growth rates. Notably, PT configurations exhibit a somewhat reversed trend, where increased $\beta_{p}$ is slightly stabilizing for the $n=0$ mode.  Elongation of inner flux surfaces for PT equilibria depends much less strongly on $\beta_p$ than it does in NT configurations, so the same destabilization mechanism that occurs in NT is minimal or nonexistent.  Stabilization with increasing $\beta_{p}$ in PT could be due to larger Shafranov shifts that displace the current density profile closer to the stabilizing vacuum vessel \cite{jin-ping_stability_nodate}.

The elongation of inner flux surfaces decreases with increasing $l_{i}$ independent of triangularity, which should lead to a improved vertical stability at higher $l_{i}$.  However, higher $l_{i}$ also corresponds to more peaked current profiles, which concentrates more of the plasma towards the core. The effective distance between the plasma and the conducting wall is subsequently decreased, so the wall has a less considerable stabilizing effect at higher values of $l_{i}$ \cite{song_impact_2021}. To generate these free-boundary equilibria with different values of $\l_\mathrm{i}$ in \texttt{TokaMaker}, the baseline profiles are modified to take the form $\frac{dp}{d\psi_\mathrm{N}} = p_0(1-\psi_\mathrm{N}^2)^\alpha$ and $\frac{1}{2}\frac{dF^2}{d\psi_\mathrm{N}} = f_0(1-\psi_\mathrm{N}^2)^\alpha$ where $\alpha$ is a free parameter that ranges from 1 to 4. Internal inductance increases with the value of $c$, enabling several scans shown in figure~\ref{fig:prof}.

For the baseline scenario presented above, vertical instability growth rates predominantly increase with increasing internal inductance in the presence of a conformal conducting wall, as shown in ~\ref{fig:profiles-b}. Unlike for poloidal beta, the relationship between internal inductance and vertical stability appears independent of triangularity. However, we note that the dependence of vertical stability on $l_\mathrm{i}$ is also influenced by plasma elongation. Figure~\ref{fig:profiles-c} shows that instability growth rates exhibit a weak negative correlation with increasing internal inductance for lower elongations.  In less elongated cross sections, the plasma is farther from the vacuum vessel on average, suggesting that wall stabilization effects play a less significant role. As a result, the increased elongation of inner flux surfaces may dominate over the redistribution of the current profile  when $\kappa$ is small.

\subsection{Aspect Ratio}
\begin{figure}
	\includegraphics[width=1\linewidth]{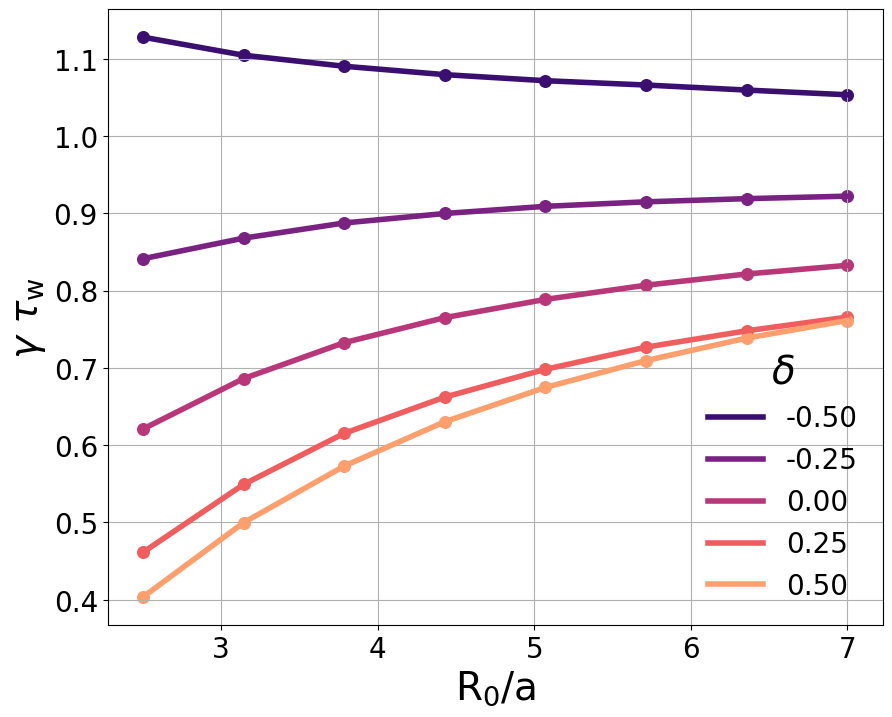}
	\caption{Feedback capability parameter as a function of device aspect ratio for equilibria with different $\delta$.}
	\label{fig:aspectratio}
\end{figure}

Another key consideration for NT reactor design is machine aspect ratio, which is expected to also impact vertical stability properties.  Aspect ratio is varied in figure \ref{fig:aspectratio} by modifying the plasma major radius while keeping the plasma minor radius fixed. For PT equilibria, growth rates are considerably lower at smaller aspect ratios, while the effect is much less pronounced for NT; growth rates decrease marginally with decreasing aspect ratio for most NT shapes.  At negative triangularities as strong as $\delta=-0.5$, growth rates actually increase with decreasing aspect ratio.

The stabilizing effect of low aspect ratios on PT plasmas is a result of increased \enquote{natural elongation} in spherical tori \cite{peng_features_1986}. At tight aspect ratios, the two sides of the plasma are in close proximity to one another, resulting in an attractive force that naturally pushes the plasma towards the inboard side without external shaping. The concentration of the plasma on the inboard side stretches the cross-section vertically.  As a result, PT plasmas in spherical tokamaks exhibit some inherent elongation without input from external shaping coils, which are responsible for the vertical instability of elongated cross-sections \cite{peng_features_1986}.  However, in NT, the elongated side of the plasma is found on the outboard side of the device, where it is not considerably \enquote{stretched} via attraction to the opposite side of the plasma.  Therefore, an NT spherical tokamak does not have the same vertical stability advantage as one might expect in the traditional PT scenario. This result challenges the applicability of the NT concept to spherical tori, though an experimental demonstration of the behavior of NT at low aspect ratio is still needed.

\section{Implications of Vacuum Vessel Shape}
\vspace{-1pt}
\label{sec:nonc}
As discussed above, NT configurations with a conformal wall are less vertically stable than traditional PT shapes, particularly at higher $\beta_p$ and lower aspect ratio. This trend is largely attributed to the increased elongation of inner flux surfaces in NT equilibria, which results purely from geometric effects. As a result, vertical stability will be a key consideration in NT reactor design. While the inherent instability of NT is a function of the equilibrium geometry, other aspects of the machine environment can be modified to improve stability.  This modification can take the form of more advanced control schemes or modifications to conducting structures in order to to improve passive stabilization mechanisms.


\subsection{Non-conformal Walls}

One method through which to improve the passive stabilization of NT configurations is to modify the relative shape of the vacuum vessel with respect to the plasma.  Because the vacuum vessel is conducting, vertical motion of the plasma induces response currents in the vacuum vessel that resist further displacement.  Traditional vacuum vessel cross-sections are designed to match the desired plasma shape in order to maximize plasma volume while limiting unused space. However, a conformal wall geometry is potentially not optimal from a vertical stability perspective because it maintains a constant wall-plasma gap, even though the plasma could respond to increased stabilization in certain key areas.

\begin{figure}
	\includegraphics[width=1\linewidth]{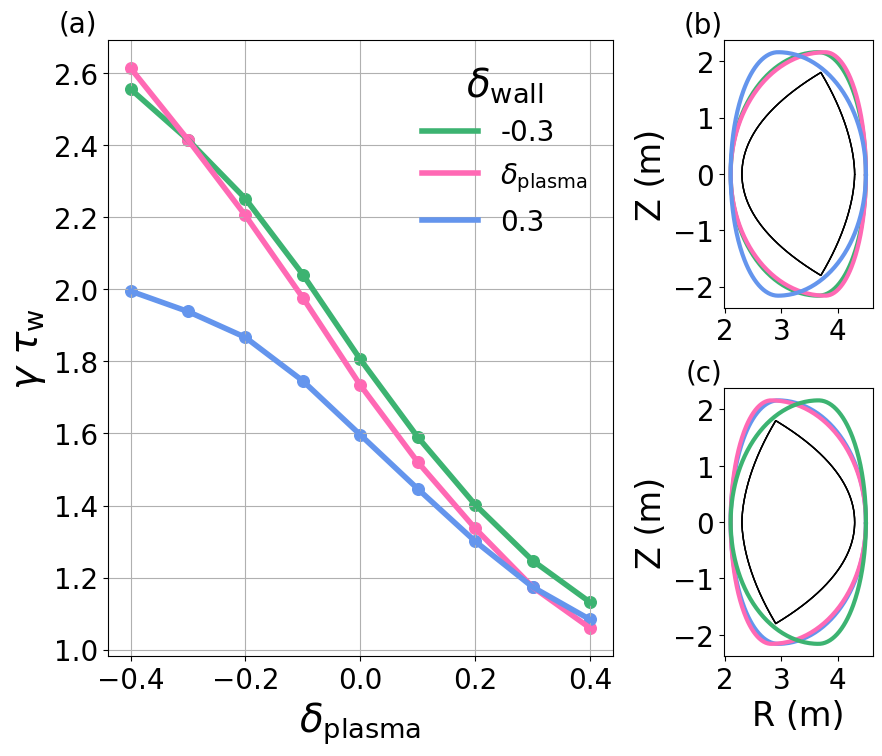}
	\caption{(a) Feedback capability parameter $\gamma \tau_w$ as a function of $\delta_{\mathrm{plasma}}$ for $\delta_{\mathrm{wall}} = -0.3$, $\delta_{\mathrm{wall}} = 0.3$ and $\delta_{\mathrm{wall}} = \delta_{\mathrm{plasma}}$ with (b) accompanying cross section for $\delta_{\mathrm{plasma}} = -0.4$. and (c) accompanying cross section for $\delta_{\mathrm{plasma}} = 0.4$}
    \label{fig:noncon}%
\end{figure}

Figure~\ref{fig:noncon} demonstrates that positive triangularity walls may stabilize NT configurations, illustrating that results from similar simulations utilizing the \texttt{AVSTAB} code \cite{song_impact_2021} persist for the free-boundary, diverted equilibria generated in \texttt{TokaMaker}.  For each value of $\delta$, three different wall shapes are considered: one with fixed positive triangularity, one with fixed negative triangularity, and one with triangularity that is modified to match the plasma triangularity.  Inner and outer midplane gaps are fixed for all wall shapes by requiring $b/a = 1.2$.
\setlength{\belowcaptionskip}{-10pt}

Wall triangularity has little effect on $\gamma \tau_{\mathrm{w}}$ for strongly PT plasma shapes.  However, NT configurations have significantly reduced growth rates when coupled with a PT vacuum vessel.  The non-conformal scenario decreases the gaps between the NT plasma shape and the vacuum vessel on the outboard side of the device, indicating that this region may be important for passive stabilization of the n=0 mode. 

Even though NT configurations have decreased vertical instability growth rates when coupled with positive triangularity walls, non-conformal walls pose engineering challenges when trying to develop a reactor concept.  For example, conformal walls provide a natural location for divertor placement without requiring significant increases in device volume because the vacuum vessel is already pushed outward at the X-points. Further, the plasma volume is limited in a non-conformal device because larger midplane gaps between the plasma and vacuum vessel are necessary in order to fit shapes with more negative $\delta$.  Because device cost scales with volume, designs that involve large amounts of wasted space are not ideal for a fusion pilot plant concept.

\subsection{Plasma "Catching" Time}

Another downside of non-conformal walls is the reduced vertical gaps between the plasma and vacuum vessel encountered when the wall and plasma triangularity differ substantially. While $\gamma \tau_{\mathrm{w}}$ is a standard metric for quantifying vertical stability, it does not provide a direct assessment of whether or not there is enough time available for control systems to \enquote{catch} the plasma during a vertical displacement event (VDE) before it becomes limited. A rough estimate for a catching time ($\tau_{\mathrm{catch}}$) can be determined using a rigid VDE assumption in which the plasma shape remains constant and the vertical position evolves as $z = z_0 e^{\gamma t}$, where $z_0$ is the height of the initial perturbation and $\gamma$ is the linear growth rate computed in \texttt{TokaMaker}, similar to the methodology used to assess vertical controllability predictions for the SPARC tokamak in reference~\cite{nelson_sparc_2024}.  

\begin{figure}%
    \labelphantom{fig:catch-a}
    \labelphantom{fig:catch-b}
    
    \centering
    \includegraphics[width=1\linewidth]{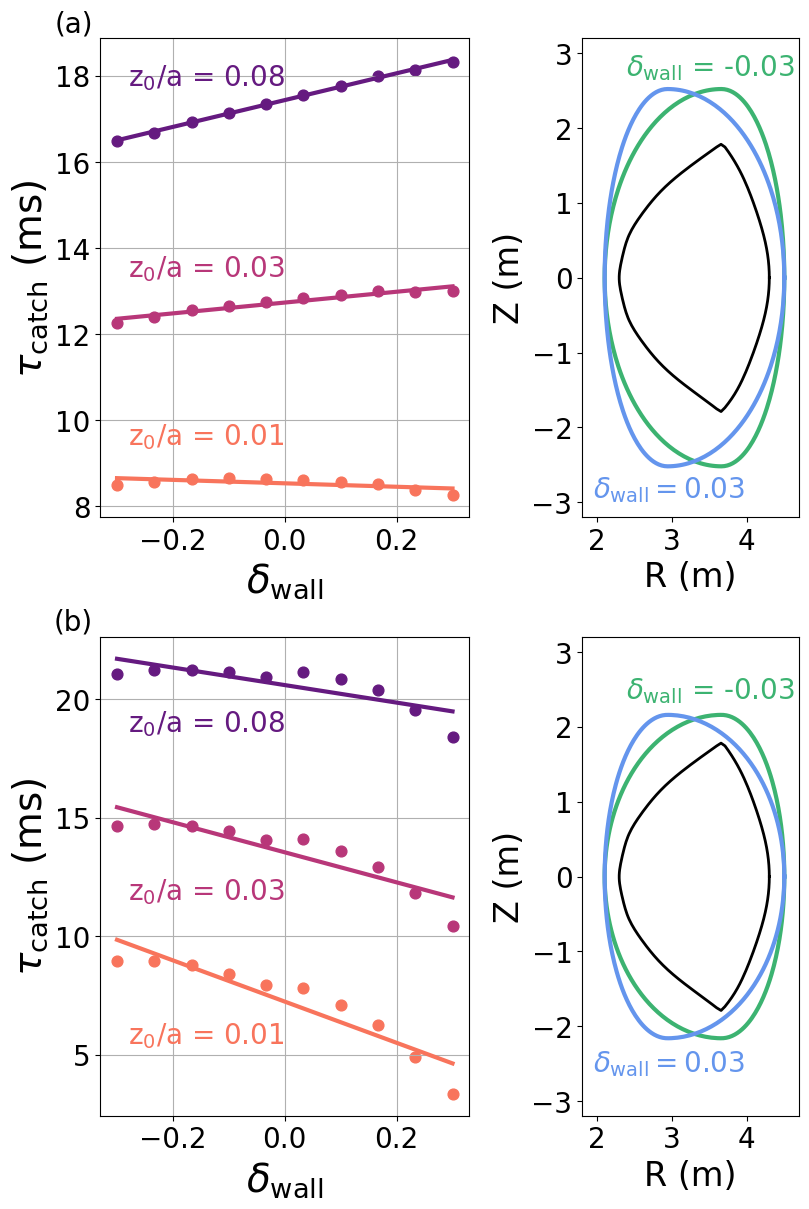} %
   
    \caption{Catching time for $\delta_{\mathrm{plasma}} = -0.4$ and  $\kappa_{\mathrm{plasma}}$ = 1.5  as a function of $\delta_{\mathrm{wall}}$ for (a) $\kappa_{\mathrm{wall}}$ = 2.1 and (b) $\kappa_{\mathrm{wall}}$ = 1.8. Accompanying cross section for $\delta_{\mathrm{wall}} = \pm 0.3$. }

    \label{fig:catch}
\end{figure}

Figure~\ref{fig:catch} shows $\tau_{\mathrm{catch}}$ as a function of wall triangularity for an NT equilibrium with $\delta = -0.4$ for three initial perturbation heights.  In figure~\ref{fig:catch-a}, $\kappa_\mathrm{plasma}=1.8$ while $\kappa_{\mathrm{wall}}=2.1$, creating larger vertical gaps. In figure~\ref{fig:catch-b}, $\kappa_{\mathrm{plasma}} = 1.8$ and $\kappa_{\mathrm{wall}} = 1.8$ in order to create the most compact device possible.  Depending on the difference in plasma and wall elongation and the scale of the initial displacement, either strongly PT or strongly NT walls can be favorable in terms of $\tau_{\mathrm{catch}}$. When $\kappa_{\mathrm{wall}}>\kappa_{\mathrm{plasma}}$, PT walls are optimal because the reduction in growth rate compensates for the slightly smaller vertical gaps.  However, when $\kappa_{\mathrm{wall}} = \kappa_{\mathrm{plasma}}$, NT walls are preferable despite the increased growth rates due to the larger vertical gaps enabled by the NT wall geometry. Because minimizing device volume is a priority, an NT wall appears to be preferable under these considerations.

The investigation of $\tau_{\mathrm{catch}}$ is multi-dimensional and the trade off between vertical gaps and reduced growth rates depends heavily on the wall diffusion times in question as well as the scale of initial perturbation assumed. For example, PT walls become more favorable for larger scale initial perturbations. Large ELMs, which are absent in NT scenarios, can trigger vertical perturbations on the order of $z_0 \approx 0.05 a$ \cite{nelson_time-dependent_2021, villone_position_2005, aliarshad_elms_1996}. More likely sources for vertical movement in NT are instabilities like sawteeth, which can trigger vertical perturbations of $z_0 \leq 0.03 a$ \cite{chapman_controlling_2010}.  For most reasonable estimates of $z_0$, the overarching trend is that the benefit conferred by non-conformal walls due to decreased growth rates is minimal in compact devices.

Even though this result indicates that conformal walls remain ideal in terms of vacuum vessel design, the study highlights a key feature of NT vertical stability. The PT vacuum vessel places conducting structures closer to the outboard corners of the plasma, pinpointing a significant region that can dictate vertical stability for NT geometries. This lends itself to the more targeted approach discussed in Section~\ref{sec:nonc}, where passive stabilizing plates are inserted at the most relevant locations along the plasma boundary.

\vspace{60pt}

\section{Stabilization of NT Plasmas with Passive Plates}
\label{sec:plate}

\vspace{-4pt}
The catching time study demonstrates that PT walls may not be the ideal solution to improve vertical stability in NT configurations due to the larger device volumes it would require to ensure control.  An alternative solution to the vertical stability problem in an NT device is to introduce passively conducting plates at locations along the plasma boundary where conductors are particularly effective. This is a relatively common approach already adopted in many modern devices such as the KSTAR \cite{kstardesign_2002} and EAST \cite{east_effect_2012} tokamaks, but has not yet been considered for NT operation.
\vspace{-11pt}
\subsection{Passive Plate Positioning}
\vspace{-4pt}
\begin{figure*}%
    \labelphantom{fig:plate-a}
    \labelphantom{fig:plate-b}
    \labelphantom{fig:plate-c}
    \labelphantom{fig:plate-d}
    \centering
    \includegraphics[width=1\linewidth]{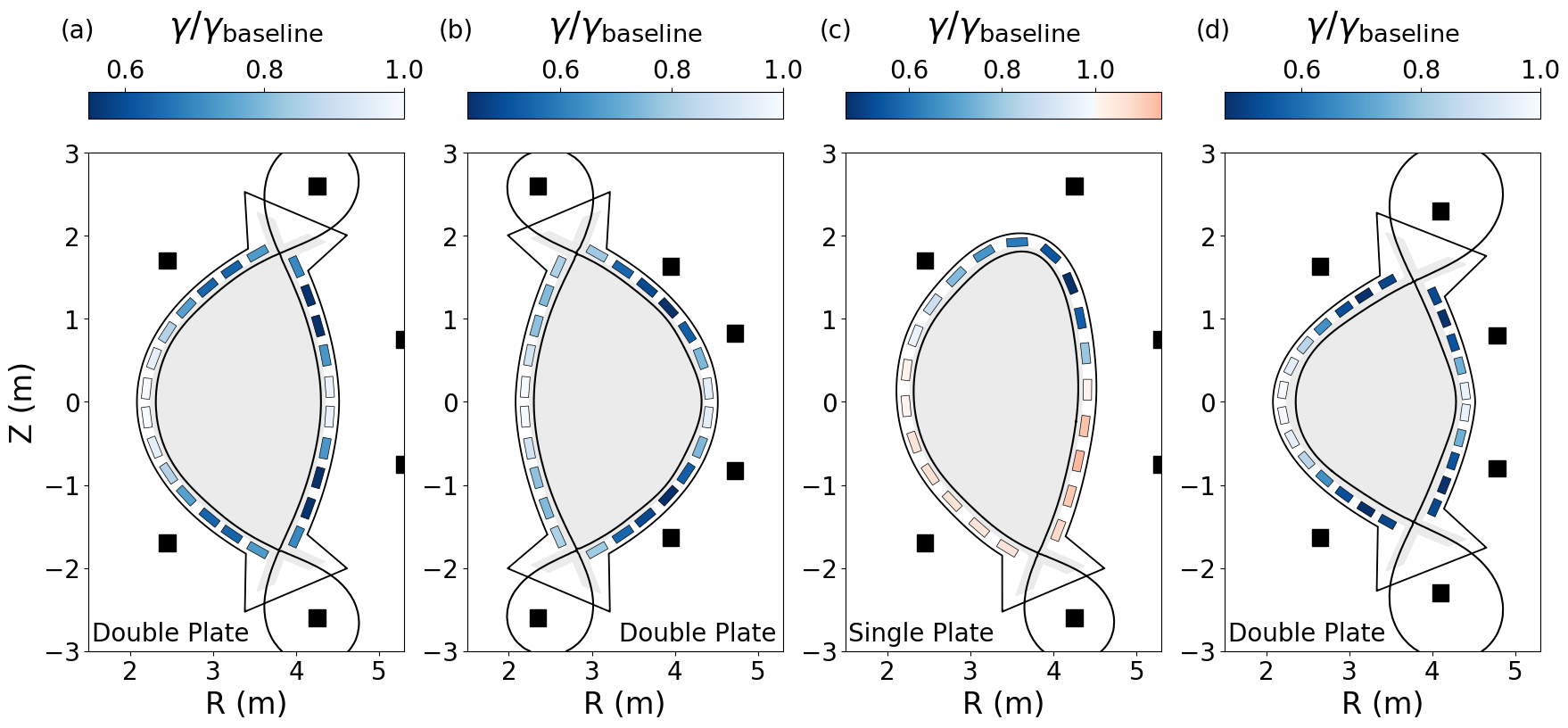} %
    \caption{ Growth rate reduction due to (a) up-down symmetric pairs of passive plates for $\delta = -0.5$ and $\kappa = 1.8$  (b) up-down symmetric pairs of passive plates for $\delta = 0.5$ and $\kappa = 1.8$ (c) individual passive plates for a SN equilibria with $\delta = 0.5$ and $\kappa = 1.8$ and (d) up-down symmetric pairs of passive plates for $\delta = -0.5$ and $\kappa = 1.5$.} %

    \label{fig:plate}%
    
\end{figure*}

The effect of passive plates is explicitly modeled in \texttt{TokaMaker} in figure~\ref{fig:plate} by introducing passive structures with high conductivity in the regions of interest and comparing the new vertical instability growth rates to the growth rate of the n=0 mode with only the vacuum vessel present. For the purpose of this study, the vacuum vessel is taken to be composed of Inconel ($\rho = 1.33\times10^{-6}\, \Omega$m) and the passive plates are taken to be copper ($\rho = 3.80\times10^{-8} \,\Omega$m) and have a cross sectional area of $20\,$$\mathrm{cm}^2$ \cite{ioki_fusion_1995}. Figure~\ref{fig:plate-a} shows the reduction in vertical instability growth rate ($\gamma/\gamma_{\mathrm{baseline}}$) due to up-down symmetric pairs of passive plates for a double-null equilibrium with $\delta = -0.5$ and $\kappa = 1.8$. The baseline equilibrium, with a conformal vacuum vessel modified to make room for realistic divertors, has $\gamma_{\mathrm{baseline}} = 95\,\mathrm{Hz}$ and $\tau_{\mathrm{w}} = 20\,\mathrm{ms}$. The greatest stabilization occurs with plates on the outboard side of the device, resulting in $\gamma/\gamma_{\mathrm{baseline}} =$ 0.56.  Passive plates on the inboard side are also shown to provide significant stabilization, reducing growth rates to $\gamma/\gamma_{\mathrm{baseline}} =$ 0.64.  

The ideal location for passive plates in NT configurations is somewhat similar to PT.  An equivalent scan of passive plate locations for a PT equilibrium in figure~\ref{fig:plate-b} similarly finds the greatest growth rate reduction for passive plates on the outboard side of the device, much like the passive plate schemes implemented in KSTAR \cite{kstardesign_2002} and EAST \cite{east_effect_2012}.  However, for PT, passive stabilizers on the inboard side have little to no impact, which is consistent with the results of figure~\ref{fig:noncon}, where walls with negative triangularity have no stabilizing effect on PT equilibria.  Therefore, NT configurations have more flexibility with respect to passive plate placement.

Figure~\ref{fig:plate-c} reveals an interesting effect that occurs when considering a single passive plate, rather than up-down symmetric pairs, and a single-null equilibrium with $\gamma_{\mathrm{baseline}} = 84\,\mathrm{Hz}$.  In this case, the most stabilizing passive plates are located near the upper outboard corner of the plasma, which is opposite the actual X-point.  Conductors located near the true X-point are found to be slightly destabilizing, increasing vertical instability growth rates to values higher than that of the baseline geometry.  

The asymmetric effect of the plates located near the upper and lower outboard corners reflects the fact that the magnetic flux cannot change drastically near the limiting point of the equilibrium \cite{hansen_2023}.  Conductors near the X-point have the effect of \enquote{squeezing} the perturbed flux towards the opposite side of the plasma, where it is more free to vary, thus destabilizing the equilibrium further. On the other hand, conductors near the opposite outboard corner are the most stabilizing because this is an area of the plasma where the flux can change considerably as a result of a perturbation, so it benefits the most from a passive stabilizing plate.  The destabilizing effect is typically mild when compared to the stabilizing effect.  The most destabilizing plates result in an $\gamma/\gamma_{\mathrm{baseline}} =$ 1.15, while the most stabilizing plates reduce growth rates to  $\gamma/\gamma_{\mathrm{baseline}} = 0.45$.

The mild destabilization that results from passive plates near the plasma X-point is not isolated to single-null equilibria, and can actually be observed in shapes that appear symmetric, such as the double-null equilibria considered in figure~\ref{fig:plate-a} and ~\ref{fig:plate-b}. Any perturbation to a symmetric equilibrium will ruin the apparent symmetry and isolate a single limiting point, so a similar conductor scan where a single passive plate is moved around the plasma boundary will be asymmetric in nature.  In the symmetric scans considered in figures~\ref{fig:plate-a} and ~\ref{fig:plate-b}, one passive plate in the pair is stabilizing while the other is actually destabilizing, but because the growth rate reduction is more significant than the growth rate enhancement, the pairs of passive plates have a net stabilizing effect. 

\setlength{\belowcaptionskip}{-10pt}
\begin{figure}[h]
    \centering
    \includegraphics[width=1\linewidth]{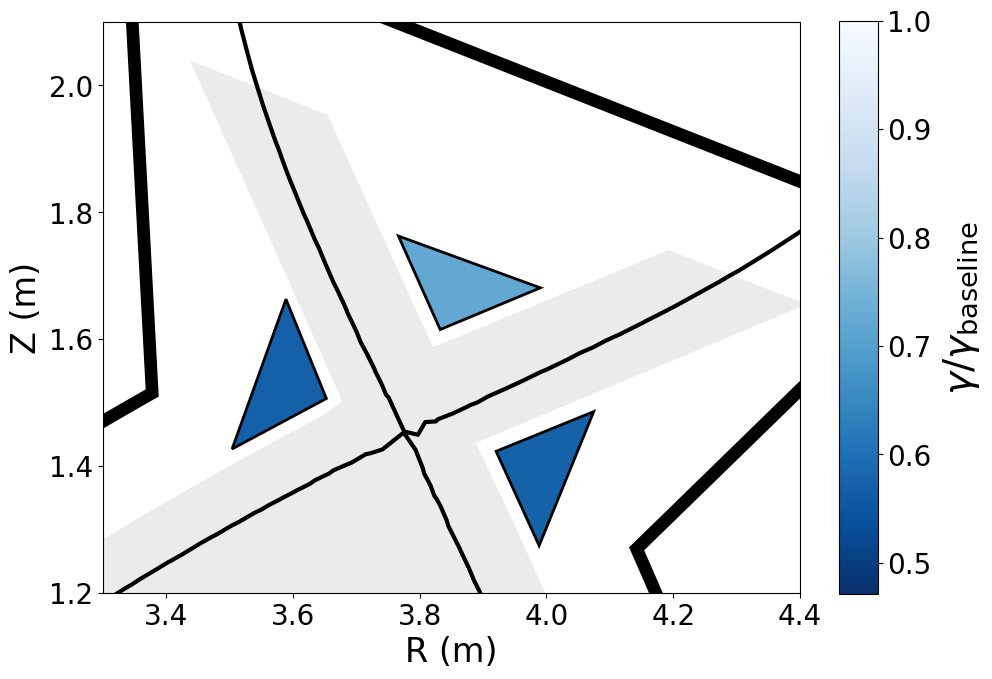} 
    \caption{Up-down symmetric pars of passive plates with triangular cross-section, referred to in the text as ``wedges", that lead to nominal growth rates of $\sim60\%$ the baseline values.} 
    \label{fig:wedge}%
\end{figure}

\vspace{24pt}
While both NT and PT equilibria benefit strongly from passive plates on the outboard side of the device, the picture for NT is slightly more complex due to the geometry of the plasma cross-section, where the X-points are also located on the outboard side. When considering an NT scenario with lower elongation, such as the one in figure ~\ref{fig:plate-d} with $\kappa = 1.5$, the locations of the X-points coincide with the regions of maximum stabilization. Because traditional passive plate geometries may be incompatible with the required divertor structures, an alternative passive plate design is considered.


\vspace{12pt}
Triangular conducting wedges with the same cross sectional area and distance from the last closed flux surface as the rectangular plates in figure~\ref{fig:plate-d} are positioned in the private flux region and adjacent to the X-points in  figure~\ref{fig:wedge}. The wedge in the private flux region is not as effective as the best rectangular passive plates at stabilizing the n=0 mode, reducing growth rates to $\gamma/\gamma_{\mathrm{baseline}} = 0.72$.  On the other hand, the wedges located adjacent to the X-points, both on the inboard and outboard side, manage to reduce growth rates to $\gamma/\gamma_{\mathrm{baseline}} = 0.57$, which is on par with the better rectangular plates. Passive plates like those in figure \ref{fig:wedge} also appear compatible with current divertor concepts.  For example, a structure similar to the the ITER divertor cassette \cite{guerrini_fabrication_2021} could incorporate stabilizers behind the inner and outer vertical targets. 

\subsection{Optimal Configurations}

The passive plate study provides an interesting alternative to larger-scale modifications of traditional vacuum vessel structures such as those considered in section~\ref{sec:nonc}. A compact device with favorable vertical stability properties might feature a conformal wall with conducting passive plates located on the outboard side for added stabilization.  One example is considered in figure~\ref{fig:design-a}, where passive plates are placed on the outboard side of the device for an NT configuration with $\kappa = 1.8$ and lead to a growth rate reduction of $\gamma/\gamma_{\mathrm{baseline}} = 0.16$.  The absolute growth rate of this scenario is 15\,Hz, which is lower than the 42\,Hz baseline growth rate of the PT equilibrium in ~\ref{fig:plate-b}. As such, inclusion of the passive plates is sufficient to make NT vertical stability comparable to, or even better than, the vertical stability in equivalent PT scenarios.  Unlike with non-conformal walls, the positioning of the plates also maintains the same vertical gap between the plasma and passive structures. Using the same approximations as in section~\ref{sec:nonc}, the inclusion of passive plates results in a six-fold improvement in $\tau_\mathrm{catch}$, increasing the likelihood of successful control system response.

\begin{figure}
    \centering
    \labelphantom{fig:design-a}
    \labelphantom{fig:design-b}
    \labelphantom{fig:design-c}
    \includegraphics[width=1\linewidth]{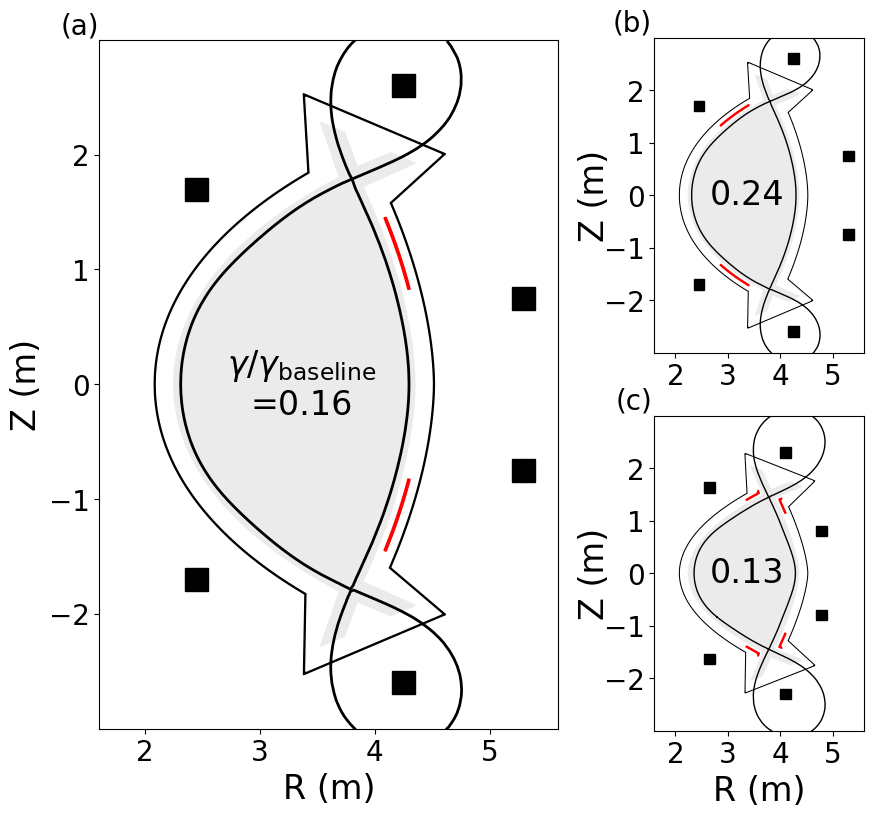}
    \caption{Potential passive plate designs for NT configurations, yielding growth rate reductions of (a) $\gamma/\gamma_{\mathrm{baseline}}$ = 0.16 for $\kappa = 1.8$, (b) $\gamma/\gamma_{\mathrm{baseline}}$ = 0.24 for $\kappa = 1.8$ and (c) $\gamma/\gamma_{\mathrm{baseline}}$ = 0.13 for $\kappa = 1.5$.} 
    \label{fig:design}%
\end{figure}


As demonstrated in ~\ref{fig:plate-a}, plates on the inboard side of the device can also lead to improved vertical stability in NT configurations. The passive plate set shown in ~\ref{fig:design-b} reduces growth rates to $\gamma/\gamma_{\mathrm{baseline}} = 0.24$ for an equilibrium with $\kappa = 1.8$.  Because vertical control coils are usually located on the outboard side of the machine, the opportunity to spatially separate active control from passive stabilizers, which reduce the response time of control systems, might be advantageous. Further time-dependent control simulations, which are briefly discussed in section~\ref{sec:conc}, would be necessary to fully explore the benefits conferred by inboard versus outboard plates.  Finally, in a device with lower target elongation, an optimal scenario may involve integration of passive stabilizers with advanced divertor concepts, such as the plates in ~\ref{fig:design-c}.  While this design confers little stability advantage for more elongated cross-sections, it is optimal for low elongation scenarios and reduces growth rates to $\gamma/\gamma_{\mathrm{baseline}} =0.13$ when $\kappa = 1.5$.

\begin{figure}
    \centering
    \includegraphics[width=1\linewidth]{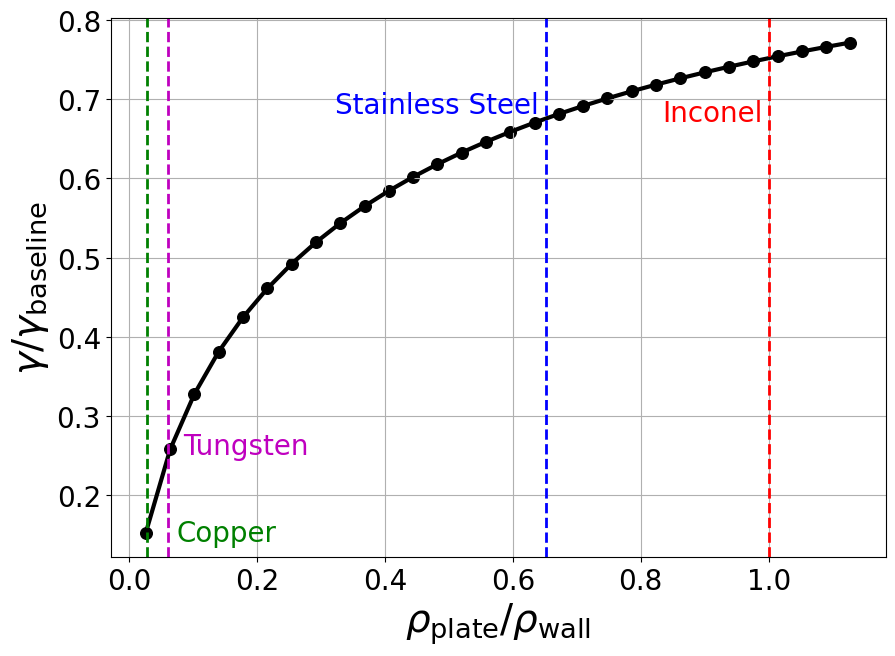}
    \caption{Growth rate reduction due to plate design in figure~\ref{fig:design-a} as a function of the plate to wall resistivity ratio ($\rho_{\mathrm{plate}}/\rho_{\mathrm{wall}}$), with common passive structure materials labeled.} 
    \label{fig:res}%
\end{figure}


The reduction in vertical instability growth rates due to passive stabilizing plates depends on the ratio of the plate resistivity to the vacuum vessel resistivity.  Figure~\ref{fig:res} shows the growth rate reduction due to the optimized plates considered in figure~\ref{fig:design-a} as a function of this ratio. It is important to note that the passive stabilizing plates only confer significant advantages when they are much more conducting than the vacuum vessel.  In addition to having low resistivity, passive plates in fusion-grade devices must also be composed of materials which are not easily activated by neutrons. Copper, despite being a common choice for passive plates in experimental tokamaks, is easily activated \cite{zhang_radioactivity_2023}.  Tungsten, due to to its relatively low resistivity and minimal activation, might be a promising alternative.

   

    


\section{Summary and Outlook}
\label{sec:conc}

Negative triangularity plasmas in conformal wall geometries are shown to be more vertically unstable than otherwise equivalent positive triangularity plasmas, which is consistent with previous work \cite{song_impact_2021,Nelson2023}. Like in PT configurations, vertical instability growth rates increase with increasing $l_\mathrm{i}$ for elongated NT cross-sections.  However, unlike PT, NT configurations are more vertically unstable at higher $\beta_p$ and are not significantly stabilized at low aspect ratios.


In section~\ref{sec:nonc}, wall shapes are shown to have a significant impact on NT vertical stability.  While a positive triangularity vacuum vessel reduces instability growth rates by decreasing the distance between the outboard side of the plasma and the conducting wall, this configuration also reduces the vertical gaps between the plasma and the top of the vessel which makes controlling VDEs more challenging in practice. By quantifying the time available to respond during a vertical displacement event, we show that, in compact devices, conformal walls are optimal for NT configurations despite the reduced instability growth rates that would be afforded by a PT vacuum vessel.

An NT pilot plant with a conformal vacuum vessel could feature passive plates for additional stabilization. The results of section~\ref{sec:plate} demonstrate that passive plates on the outboard side provide the greatest reduction in vertical instability growth rates. An optimized device for a strongly NT and high $\kappa$ scenario features passive plates on the outboard side, which were found to reduce nominal growth rates to 16\% of their baseline value, making the configuration more stable than the equivalent PT equilibrium without plates.  Unlike for PT configurations, NT equilibria are also stabilized by passive plates on the inboard side, providing greater flexibility for integration with active control coils and other engineering constraints.  For lower target elongations, the region of maximum stabilization coincides with the X-points.  In this scenario, passive plates with triangular cross section adjacent to the X-points were found to be a potential alternative to traditional passive plate designs which could be easily incorporated into advanced divertors.  This study shows that informed placement of conducting structures can significantly improve the vertical stability landscape for NT without sacrificing any inherent properties of the NT core or divertor scenario and without requiring extensive excess vacuum vessel volume. 

 The interplay between passive stabilizers and active control systems is multifaceted and warrants further investigation. By slowing down the plasma motion, the addition of passive plates can reduce the effectiveness of active coils in quickly changing the fields to respond to VDEs. However, passive plates can also reduce the induced current in the active coils that must be mitigated by the power systems during a disruption and can also reduce the voltage requirements on the active coil required for vertical stability, particularly if the passive plates are poloidally separated from the vertical stability coils. To fully understand the potential benefits and interactions between a passive plate set and active control schemes, as in reference~\cite{nelson_sparc_2024}, a full model of the target device is needed, including accurate descriptions of the power supplies and control algorithms.  Furthermore, modeling will be necessary to predict the forces experienced by different passive plate designs during disruption events \cite{pustovitov_models_2022}.

This level of description is outside the scope of the present work. Future studies should involve leveraging the time dependent simulation capabilities available in \texttt{TokaMaker} or other codes to study active control systems and advanced feedback algorithms by modifying coil currents in response to changes in plasma position.  A time-dependent simulation would provide more robust quantification of vertical control requirements for an NT reactor and allow for the optimization of coil sets with strong vertical control capabilities. These studies could include the effects of noise and the magnitude of initial perturbation to recover \cite{humphreys_experimental_2009}. Further NT vertical control experiments, such as those presented in \cite{Nelson2023}, are also desirable in order to validate theoretical vertical instability calculations for a wider range of plasma geometries and conditions. 

\section*{Acknowledgements}
This work was supported by the U.S. Department of Energy, Office of
Science, Office of Fusion Energy Sciences under Award(s) DE-SC0022270, DE-SC0019239,
DE-SC0019479, DE-SC0022272. S. Guizzo was supported by the U.S. Department of Energy Science Undergraduate Laboratory Internship (SULI) program and Columbia University internal funds.  Cross-code verification on the SPARC tokamak was supported by Commonwealth Fusion Systems.
\label{sec:ack}

\section*{Appendix A: TokaMaker Benchmark}
\label{sec:val}

Because this is the first study utilizing \texttt{TokaMaker} vertical instability growth rate calculations, a benchmark with the established \texttt{TokSys} code \cite{humphreys_development_2007} is conducted. The \texttt{GSPert} module of the \texttt{TokSys} suite can be used to compute the open loop vertical instability growth rates for equilibria in well-defined machine geometries.

\begin{figure}
    \centering
    \includegraphics[width=1\linewidth]{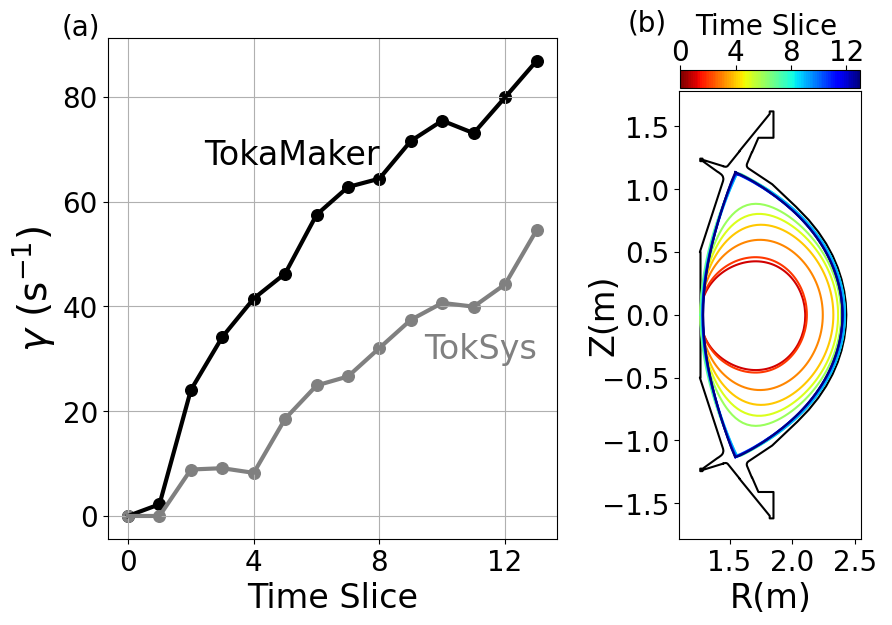}
    \caption{a) Vertical instability growth rates calculated with the \texttt{TokaMaker} and \texttt{TokSys} codes for equilibria at time slices along a SPARC discharge trajectory and b) plasma boundary at each time slice.} 
    \label{fig:val}%
\end{figure}

For this comparison, a time series of equilibria along a potential SPARC discharge trajectory are recreated in both \texttt{TokSys} and \texttt{TokaMaker} and the n=0 instability growth rates are computed.  As shown in figure~\ref{fig:val}, the growth rates show close qualitative agreement throughout the entire trajectory as both the geometry and equilibrium profiles evolve. Because this study involves assessment of trends for vertical stability in NT reactor scenarios, the qualitative agreement with the \texttt{TokSys} code is deemed sufficient to validate the \texttt{TokaMaker} calculations.

The absolute magnitude of growth rates computed in \texttt{TokaMaker} are greater than those computed in \texttt{TokSys}.  The discrepancy could be due to differing representations of induced currents in passive conducting structures. In \texttt{TokSys} the induced currents are uniform across the conductor width, while \texttt{TokaMaker} evolves the currents in the passive structures, allowing for spatially nonuniform current distributions. Because the walls of SPARC are relatively thick, this difference in representation could have significant impacts on growth rate magnitudes.  Furthermore, small modifications to the vacuum vessel implemented in the \texttt{TokSys} SPARC model are necessary in the conversion to the \texttt{TokaMaker} geometry description, providing another potential explanation for discrepancies in absolute growth rates.

\end{document}